\documentclass[final,3p,times]{elsarticle}

\usepackage{lineno,hyperref}
\usepackage{amssymb}
\usepackage{graphicx}
\usepackage{amsmath}
\usepackage{amsfonts,color}
\usepackage{amsfonts}
\usepackage{float}
\usepackage{longtable}
\providecommand{\U}[1]{\protect\rule{.1in}{.1in}}
\modulolinenumbers[99]
\begin{document}

\begin{abstract}
\noindent This paper deals with the time differential dual-phase-lag heat
transfer models aiming, at first, to identify the eventually restrictions that make them
thermodynamically consistent. At a first glance it can be observed that the capability of a
time differential dual-phase-lag model of heat conduction to describe real phenomena depends
on the properties of the differential operators involved in the related constitutive equation. In fact,
the constitutive equation is viewed as an ordinary differential equation in terms of the heat
flux components (or in terms of the temperature gradient) and it results that,
for approximation orders greater than or equal to five, the corresponding characteristic equation has at least
a complex root having a positive real part. That leads to a heat flux component (or temperature gradient) that grows to infinity when the time tends to infinity and so there occur some instabilities.  Instead, when the approximation orders
are lower than or equal to four, this is not the case and there is the need to study the compatibility with the Second Law of Thermodynamics.
To this aim the related constitutive equation is reformulated within the system of the fading memory theory, and thus the heat flux vector is written in terms of the history of the temperature gradient and on this basis the compatibility of the model with the
thermodynamical principles is analyzed.

Link to publisher: \url{https://doi.org/10.1016/j.ijheatmasstransfer.2017.06.071}
\end{abstract}

\begin{frontmatter}

\title{On the thermomechanical consistency of the time differential dual-phase-lag models of heat conduction}

\author[Chirita1,Chirita2]{Stan Chiri\c{t}\u{a}\corref{mycorrespondingauthor}}
\cortext[mycorrespondingauthor]{Corresponding author}
\address[Chirita1]{\small Faculty of Mathematics, Al. I. Cuza University of Ia\c{s}i - 700506 Ia\c{s}i, Romania \vspace{0.5mm}}
\address[Chirita2]{\small Octav Mayer Mathematics Institute, Romanian Academy - 700505 Ia\c si, Romania\vspace{2.5mm}}
\ead{schirita@uaic.ro}

\author[CiarlettaTibullo]{Michele Ciarletta}
\address[CiarlettaTibullo]{\small University of Salerno, via Giovanni Paolo II n. 132 - 84084 Fisciano (SA), Italy }
\ead{mciarletta@unisa.it}

\author[CiarlettaTibullo]{Vincenzo Tibullo}
\ead{vtibullo@unisa.it}


\end{frontmatter}

\section{Introduction}

The dual-phase-lag model of heat conduction proposed in \cite{Tzou1995a}-\cite{Tzou1995c} distinguishes the time instant $t + \tau_q$, at which the
heat flux flows through a material volume and the time instant $t + \tau_T$, at which the temperature gradient establishes across
the same material volume:
\begin{equation}
q_{i}(\mathbf{x},t+\tau_{q})=-k_{ij}(\mathbf{x})T_{,j}(\mathbf{x},t+\tau
_{T}),\qquad\text{with }\quad\tau_{q},\, \tau_{T}\geq 0. \label{i1}%
\end{equation}
The above constitutive equation states, synthesizing its meaning, that the temperature gradient $T_{,j}$\ at
a certain time $t+\tau_{T}$ results in a heat flux vector $q_{i}$\ at a
different time $t+\tau_{q}$. In the above constitutive equation (\ref{i1}),
besides the explicit dependence upon the spatial variable, we point out that
$q_{i}$ are the components of the heat flux vector, $T$ represents the
temperature variation from the constant reference temperature $T_{0}>0$ and
$k_{ij}$ are the components of the conductivity tensor; moreover, $t$ is the
time variable while $\tau_{q}$ and $\tau_{T}$ are the phase lags (or delay
times) of the heat flux and of the temperature gradient, respectively. In
particular, $\tau_{q}$ is a relaxation time connected to the fast-transient
effects of thermal inertia, while $\tau_{T}$ is caused by microstructural
interactions, such as phonon scattering or phonon-electron interactions
\cite{TzouBook}.
In addition to the thermal conductivity, the phase lags $\tau_T$ and
$\tau_q$ are treated as two additional intrinsic thermal properties
characterizing the energy-bearing capacity of the material.

Equation (\ref{i1}) describing the lagging behavior in
heat transport, when coupled with the energy equation
\begin{equation}
-q_{i,i}+\varrho r=a\frac{\partial T}{\partial t},\label{i2}
\end{equation}
displays two coupled differential equations of
a delayed type. Due to the general time shifts at different
scales, $\tau_T$ and $\tau_q$, no general solution has been known yet.
The refined structure of the lagging response depicted by equations
(\ref{i1}) and (\ref{i2}), however, has been illustrated by Tzou \cite{Tzou1995c} by expanding equation (\ref{i1})
in terms of the Taylor's series with respect to time:
\begin{equation}
\begin{split}
&q_i\left(\mathbf{x},t\right)+\frac{\tau_q}{1!}\, \frac{\partial q_i}{\partial t}\left(\mathbf{x},t\right)+\frac{\tau_q^2}{2!}\, \frac{\partial^2 q_i}{\partial t^2}\left(\mathbf{x},t\right)+...+\frac{\tau_q^n}{n!}\, \frac{\partial^n q_i}{\partial t^n}\left(\mathbf{x},t\right)\\
&=-k_{ij}(\mathbf{x})\left[T_{,j}\left(\mathbf{x},t\right)+\frac{\tau_T}{1!}\, \frac{\partial T_{,j}}{\partial t}\left(\mathbf{x},t\right)+\frac{\tau_T^2}{2!}\, \frac{\partial^2 T_{,j}}{\partial t^2}\left(\mathbf{x},t\right)+...+\frac{\tau_T^m}{m!}\, \frac{\partial^m T_{,j}}{\partial t^m}\left(\mathbf{x},t\right)\right].\label{i3}
\end{split}
\end{equation}
An interesting discussion concerning this expansion has been developed by Tzou \cite{Tzou1995c} when $n-m=0$ or $n-m=1$, relating the progressive interchange between the diffusive and wave behaviors.

We emphasize that the related time differential models obtained considering
the Taylor series expansions of both sides of the equation (\ref{i1}) and retaining
terms up to suitable orders in $\tau_{q}$ and $\tau_{T}$ (namely, first or second orders in $\tau_{q}$ and $\tau_{T}$) have been widely
investigated with respect to their thermodynamic consistency as well as to
interesting stability issues (see, for example, \cite{FabrizioJTS},
\cite{FabrizioIJHMT} and \cite{Quintanilla2002}). However, the general form of the time differential dual-phase-lag model as given by (\ref{i3}) wasn't treated up to now, except for the paper by Quintanilla and Racke \cite{Quintanilla2015}, where the spatial behavior is studied for solutions of the equation obtained
by eliminating the heat flux vector between the constitutive equation (\ref{i3}) and the energy equation (\ref{i2}), provided $n=m$ or $n=m+1$.

The main purpose of this paper is to study the thermodynamical and mechanical consistency of the constitutive equation (\ref{i3}). We infer that the feasibility study of this constitutive equation greatly depends on the structure of the differential operators involved in its mathematical expression. In fact, if we consider the constitutive equation (\ref{i3}) as an ordinary linear differential equation in terms of the unknown function $q_i(t)$ (or, equivalently, in terms of the unknown function $T_{,i}(t)$) then we can observe that for $n\geq 5$ (or $m\geq 5$) it admits at least a complex root having a positive real part. That implies that $q_i(t)$ (or $T_{,i}(t)$) can tends to infinity when the time tends to infinity and so we are led to instability situations. On this way we conclude that the time differential dual-phase-lag model based on a constitutive equation of type (\ref{i3}) with $n\geq 5$ or $m\geq 5$ cannot be considered able to describe real mechanical situations. Instead, when $n=0,1,2,3,4$ and $m=0,1,2,3,4$ this is not the case and we have to study the thermodynamic consistency of the corresponding model. To this aim we follow \cite{FabrizioIJHMT} and \cite{Chirita2016} and we reformulate the constitutive equation (\ref{i3}) in such a way that the heat flux vector $q_{i}$ depends on the history of the temperature gradient. In this sense we rewrite the equation (\ref{i3}) in the framework of
Gurtin-Pipkin \cite{GurtinPipkin} and Coleman-Gurtin \cite{Coleman} fading
memory theory, and on this basis we analyze the compatibility of the model
with the thermodynamical principles. Precisely, the thermodynamic consistency of the model in concern is established when $(m,n)\in \{(0,0), (1,0), (0,1), (2,1), (1,2), (2,2), (3,2), (2,3), (3,3), (3,4), (4,3), (4,4)\}$, provided appropriate restrictions are placed on the delay times.

\section{Thermomechanical consistency of the model}

In this Section we consider the equation (\ref{i3}) as an ordinary linear non-homogeneous differential (in time variable) equation in terms of the heat flux vector components and observe that its characteristic equation is
\begin{equation}
\frac{1}{n!}\, \tau_q^n\, \lambda^n+\frac{1}{(n-1)!}\, \tau_q^{n-1}\, \lambda^{n-1}+...+\frac{1}{2!}\, \tau_q^2\, \lambda^2+\frac{1}{1!}\, \tau_q\, \lambda+1=0.\label{t1}
\end{equation}
This equation is connected with the partial sums of the Maclaurin series for the exponential function $e^z$ and with the incomplete gamma function and its roots have been intensively studied in literature (see e. g. Enestr\"{o}m \cite{Enestrom}, \cite{Enestrom1920}, Kakeya \cite{Kakeya}, Marden \cite{Marden}). On the basis of the Enestr\"{o}m-Kakeya theorem it follows that all the roots of the equation (\ref{t1}) lie outside of the disk of radius $\frac{1}{\tau_q} $. Moreover, the equation (\ref{t1}) has no real root if $n$ is even, while when $n$ is odd, it has only one real root. However, here we are interested if this equation has at least a complex root with a positive real part. To this aim we outline the results obtained by G\'{a}bor Szeg\"{o} \cite{Szego} and Jean Dieudonn\'{e} \cite{Dieudonne} who showed that the roots of the scaled exponential sum function approach the portion of the  Szeg\"{o} curve: $|z\exp(1-z)|=1$ within the unit disk as $n\rightarrow \infty$. Moreover, with the aim to visualize this result for $n\geq 5$ we recommend the simulation for the Wolfram Mathematica 11 presented in the Appendix (see also the Fig. 1). 

\begin{figure}
\includegraphics[width=\textwidth]{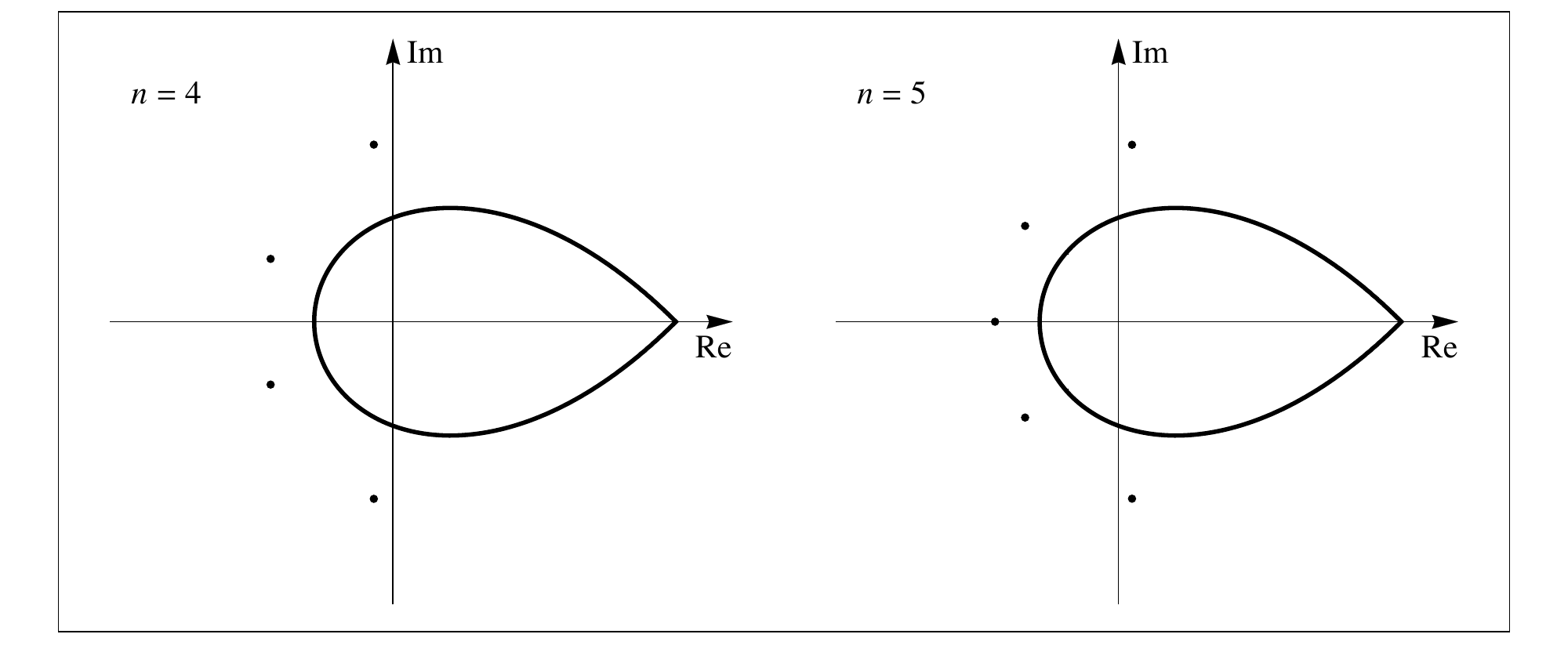}
\caption{Roots of the exponential sum for $n=4$ and forn $n=5.$}
\end{figure}

In view of the above discussion we can conclude that, for values of $n$ greater than or equal to five, the characteristic equation (\ref{t1}) admits at least one complex root having a positive real part and this leads to a couple $\{T,q_i\}$ going to infinity as time tends to infinity, that is we are led to instability of the system. In conclusion, for $n\geq 5$ the corresponding models of dual-phase-lag of heat conduction are not suitable to describe the lagging behavior.

Furthermore, we write the constitutive equation (\ref{i3}) in the following form
\begin{equation}
\begin{split}
& T_{,i}\left(\mathbf{x},t\right)+\frac{\tau_T}{1!}\, \frac{\partial T_{,i}}{\partial t}\left(\mathbf{x},t\right)+\frac{\tau_T^2}{2!}\,  \frac{\partial^2 T_{,i}}{\partial t^2}\left(\mathbf{x},t\right)+...+\frac{\tau_T^m}{m!}\, \frac{\partial^m T_{,i}}{\partial t^m}\left(\mathbf{x},t\right)\\
&=-K_{ij}(\mathbf{x})\left[q_j\left(\mathbf{x},t\right)+\frac{\tau_q}{1!}\, \frac{\partial q_j}{\partial t}\left(\mathbf{x},t\right)+\frac{\tau_q^2}{2!}\,  \frac{\partial^2 q_j}{\partial t^2}\left(\mathbf{x},t\right)+...+\frac{\tau_q^n}{n!}\, \frac{\partial^n q_j}{\partial t^n}\left(\mathbf{x},t\right)\right],\label{t2}
\end{split}
\end{equation}
where $K_{ij}$ are so that
\begin{equation}
K_{ij}k_{js}=k_{ij}K_{js}=\delta_{is}.\label{t3}
\end{equation}
Following an argument similar to that in the above discussion, this time the unknown function being considered  $T_{,i}$, we can conclude that, for values of $m$ greater than or equal to five, the corresponding constitutive equation (\ref{i3}) leads to models of dual-phase-lag of heat conduction that are not suitable to describe the lagging behavior.

On the other side, as it can be seen from the Table 1, it follows that for $n=1,2,3,4$ all the solutions of the equation (\ref{t1}) have a negative real part and hence the corresponding contribution in the heat flux vector components can lead to an asymptotic stable behavior in time couple $\{T,q_i\}$. In such cases we have to study further the consistency of the constitutive equation (\ref{i3}) with the Second Law of Thermodynamics.

\begin{center}
{\footnotesize Table 1: The values of $x=\tau_q\lambda$ with $\lambda$ solution of the characteristic equation (\ref{t1}) for $n=1,2,3,4$}
\end{center}

\hspace*{-1cm}
\begin{center}
\begin{tabular}{|c|c|c|c|c|}
\hline
$n$ & $x=\tau_q\lambda$   \\
\hline
1& $-1$  \\
 \hline
2& $-1.0 \pm 1.0\, i$   \\
 \hline
3& $-1.5961 $\\
 \text{}& $-0.70196 \pm 1.8073\, i$ \\
  \hline
4& $-0.27056 \pm 2.5048\, i$\\
 \text{}& $-1.7294 \pm 0.88897\, i$ \\
 \hline
\end{tabular}
\end{center}

In conclusion, the constitutive equation (\ref{i3}) can be thermodynamically consistent only for $n\in \{0,1,2,3,4\}$ and $m\in \{0,1,2,3,4\}$. Thus, in what follows we will study
the thermodynamic consistency of the constitutive equation (\ref{i3}) for $(n,m)\in \{0,1,2,3,4\}\times\{0,1,2,3,4\}$.

For future convenience, we will discuss separately each of the above cases. In order to study the cases involved in the Table 1, we consider the constitutive equation (\ref{i3}) as a memory constitutive equation of the type described in Gurtin-Pipkin \cite{GurtinPipkin} and Coleman-Gurtin \cite{Coleman}.

\section{Case $(n,m)=(0,0)$}
This case yields the classical Fourier law of heat flux which is compatible with the Second Law of Thermodynamics.

\section{Case $(n,m)\in \{(1,0),(0,1)\}$}
The case $(n,m)=(1,0)$ yields the Cattaneo-Maxwell equation of the heat flux vector which is consistent with thermodynamics for all $\tau_q>0$ and it is equivalent to the constitutive equation with fading memory
\begin{equation}
q_i(\mathbf{x},t)=-\frac{1}{\tau_q}\int_0^{\infty}e^{-s/\tau_q}k_{ij}(\mathbf{x})T_{,j}(\mathbf{x},t-s)ds.\label{t4}
\end{equation}
The case $(n,m)=(0,1)$ can be treated by a similar way in view of the form (\ref{t2}) of the constitutive equation and we will have
\begin{equation}
T_{,i}(\mathbf{x},t)=-\frac{1}{\tau_T}\int_0^{\infty}e^{-s/\tau_T}K_{ij}(\mathbf{x})q_{j}(\mathbf{x},t-s)ds.\label{t5}
\end{equation}
Hence we conclude that the corresponding model is thermodynamically consistent for all $\tau_T>0$.

\section{Case $(n,m)=(1,1)$}
The constitutive equation (\ref{i3}), for $(n,m)=(1,1)$, is equivalent to the constitutive equation with fading memory
\begin{equation}
q_i(\mathbf{x},t)=-\frac{1}{\tau_q}\int_0^{\infty}e^{-s/\tau_q}k_{ij}(\mathbf{x})\left[T_{,j}(\mathbf{x},t-s)+\tau_T\dot T_{,j}(\mathbf{x},t-s)\right]ds,\label{t6}
\end{equation}
which integrated by parts gives
\begin{equation}
q_i(\mathbf{x},t)=-\frac{\tau_T}{\tau_q}k_{ij}(\mathbf{x})T_{,j}(\mathbf{x},t)-\frac{1}
{\tau_q}\left(1-\frac{\tau_T}{\tau_q}\right)\int_0^{\infty}e^{-s/\tau_q}
k_{ij}(\mathbf{x})T_{,j}(\mathbf{x},t-s)ds.\label{t7}
\end{equation}

To determine the restrictions imposed by thermodynamics on
the constitutive equation in concern, we postulate the Second Law of Thermodynamics
in terms of a Clausius-Duhem inequality formulated on cyclic histories, that is (see, e.g. Amendola et al. \cite{Amendola2012}):
\begin{equation}
\oint q_i(t)T_{,i}(t)dt\leq 0,\label{t8}
\end{equation}
where the equality occurs only for the null cycle and having omitted everywhere the explicit dependence upon the space variable. Consequently, any cycle characterized by the history
\begin{equation}
T_{,i}(t-s)=f_i\cos\omega (t-s)+g_i\sin\omega (t-s), \ \omega>0, \ f_if_i+g_ig_i>0,\label{t9}
\end{equation}
has to fulfil (\ref{t8}) as an inequality.

In view of the relations (\ref{t7}) and (\ref{t9}), we have
\begin{equation}
\begin{split}
q_i(t)T_{,i}(t)&=-\frac{\tau_T}{2\tau_q}\left[k_{ij}f_if_j+k_{ij}g_ig_j+\left(k_{ij}f_if_j-k_{ij}g_ig_j\right)\cos 2\omega t+2k_{ij}f_ig_j\sin 2\omega t\right]\\
&-\frac{1}{\tau_q}\left(1-\frac{\tau_T}{\tau_q}\right)\int_0^{\infty}e^{-s/\tau_q}\left[\frac12 \left(k_{ij}f_if_j+k_{ij}g_ig_j\right)\cos \omega s\right.\\
&\left.+\frac12\left(k_{ij}f_if_j-k_{ij}g_ig_j\right)\cos \omega\left(2t-s\right)+k_{ij}f_ig_j\sin \omega\left(2t-s\right)\right]ds,\label{t10}
\end{split}
\end{equation}
so that, by replacing into the inequality (\ref{t8}), we get
\begin{equation}
\begin{split}
\int_0^{2\pi/\omega} q_i(t)T_{,i}(t)dt=-\frac{\pi\left(k_{ij}f_if_j+k_{ij}g_ig_j\right)}{\omega\left(1+\tau_q^2\omega^2\right)}
\left(\tau_T\tau_q\omega^2+1\right),\label{t11}
\end{split}
\end{equation}
which is negative for all $\omega>0$.

Concluding, the dual-phase-lag model of heat conduction for $(n,m)=(1,1)$ is compatible with thermodynamics for all $\tau_q>0$ and $\tau_T>0$.

\section{Case $(n,m)\in \{(2,0),(0,2)\}$}
Let us first consider the case $(n,m)=(2,0)$, that is we study the constitutive equation
\begin{equation}
q_i(t)+\tau_q \dot q_i(t)+\frac12\, \tau_q^2\ddot q_i(t)=-k_{ij}T_{,j}(t),\label{t12}
\end{equation}
where a superposed dot denotes the time differentiation. Then it can be written in the fading memory theory as
\begin{equation}
q_i(t)=-\frac{2}{\tau_q}\int_0^{\infty}e^{-s/\tau_q}\left(\sin \frac{s}{\tau_q}\right)\, k_{ij}T_{,j}(t-s)ds,\label{t13}
\end{equation}
and hence, by taking into account the relation (\ref{t9}), we have
\begin{equation}
\begin{split}
q_i(t)T_{,i}(t)&=-\frac{1}{\tau_q}\int_0^{\infty}e^{-s/\tau_q}\left(\sin \frac{s}{\tau_q}\right)\left[ \left(k_{ij}f_if_j+k_{ij}g_ig_j\right)\cos \omega s\right.\\
&\left.+ \left(k_{ij}f_if_j-k_{ij}g_ig_j\right)\cos \omega\left(2t-s\right)+2k_{ij}f_ig_j\sin \omega\left(2t-s\right)\right]ds.\label{t14}
\end{split}
\end{equation}
Therefore, we obtain
\begin{equation}
\begin{split}
&\int_0^{2\pi/\omega} q_i(t)T_{,i}(t)dt=-\frac{2\pi}{\tau_q\omega}\left(k_{ij}f_if_j+k_{ij}g_ig_j\right)\int_0^{\infty}e^{-s/\tau_q}\left(\sin \frac{s}{\tau_q}\right)\cos\omega sds\\
&=-\frac{2\pi\left(k_{ij}f_if_j+k_{ij}g_ig_j\right)}{\omega\left(\tau_q^4\omega^4+4\right)}
\left(2-\tau_q^2\omega^2\right),\label{t15}
\end{split}
\end{equation}
which cannot conserve a constant sign for all $\omega>0$. Thus, the Second Law of Thermodynamics cannot be satisfied and the corresponding dual-phase-lag model is inconsistently from thermodynamic point of view.

In view of the form (\ref{t2}) of the constitutive equation and by using an argument similar with that into above we conclude that the corresponding model for $(n,m)=(0,2)$ is incompatible with thermodynamics.

\section{Case $(n,m)\in \{(2,1),(1,2)\}$}
We outline here that the case $(n,m)=(2,1)$ was studied by Fabrizio and Lazzari in \cite{FabrizioIJHMT} and it was shown that the corresponding model, based on the constitutive equation
\begin{equation}
q_i(t)+\tau_q \dot q_i(t)+\frac12\, \tau_q^2\ddot q_i(t)=-k_{ij}\left(T_{,j}(t)+\tau_T \dot T_{,j}(t)\right),\label{t16}
\end{equation}
is compatible with the thermodynamics if and only if the delay times satisfy the inequality
\begin{equation}
0<\tau_q\leq 2\tau_T.\label{t17}
\end{equation}

In what follows we consider the case $(n,m)=(1,2)$, that is we study the constitutive equation
\begin{equation}
q_i(t)+\tau_q\dot q_i(t)=-k_{ij}\left[T_{,j}(t)+\tau_T\dot T_{,j}(t)+\frac12\, \tau_T^2\ddot T_{,j}(t)\right].\label{t18}
\end{equation}
Considering (\ref{t18}) as a differential equation with the unknown function $q_i$, we obtain the following equivalent representation
\begin{equation}
q_i(t)=-\frac{1}{\tau_q}\int_0^{\infty}e^{-s/\tau_q}k_{ij}\left[T_{,j}(t-s)+\tau_T\dot T_{,j}(t-s)+\frac12\, \tau_T^2\ddot T_{,j}(t-s)\right]ds,\label{t19}
\end{equation}
and furthermore, by successive integration by parts, we have
\begin{equation}
\begin{split}
q_i(t)=&-\frac{\tau_T^2}{2\tau_q}\, k_{ij}\dot T_{,j}(t)-\frac{\tau_T}{\tau_q}\left(1-\frac{\tau_T}{2\tau_q}\right)
k_{ij}T_{,j}(t)\\
&-\frac{1}{\tau_q}\left[1-\frac{\tau_T}{\tau_q}\left(1-\frac{\tau_T}{2\tau_q}\right)\right]
\int_0^{\infty}e^{-s/\tau_q}k_{ij}T_{,j}(t-s)ds.\label{t20}
\end{split}
\end{equation}
Thus, with the choice (\ref{t9}), from (\ref{t20}) we deduce
\begin{equation}
\int_0^{2\pi/\omega} q_i(t)T_{,i}(t)dt=-\frac{\pi\left(k_{ij}f_if_j+k_{ij}g_ig_j\right)}{\omega\left(1+\tau_q^2\omega^2\right)}
\left[\tau_T\tau_q\left(1-\frac{\tau_T}{2\tau_q}\right)\omega^2+1\right],\label{t21}
\end{equation}
and it will be negative for all $\omega>0$ if and only if the following inequality
\begin{equation}
0<\tau_T\leq 2\tau_q,\label{t22}
\end{equation}
is fulfilled.

Concluding, the dual-phase-lag models based upon the constitutive equations (\ref{t16}) and (\ref{t18}) are compatible with the thermodynamics, provided the corresponding inequalities (\ref{t17}) and (\ref{t22}) are fulfilled.

\section{Case $(n,m)=(2,2)$}

The restrictions upon the delay times which follow from Second Law of Thermodynamics for the constitutive equation
\begin{equation}
q_i(t)+\tau_q \dot q_i(t)+\frac12\, \tau_q^2\ddot q_i(t)=-k_{ij}\left(T_{,j}(t)+\tau_T \dot T_{,j}(t)+\frac12\, \tau_T^2\ddot T_{,j}(t)\right),\label{t23}
\end{equation}
have been established by Fabrizio and Lazzari in \cite{FabrizioIJHMT} to be
\begin{equation}
\left(2-\sqrt{3}\right)\tau_T<\tau_q<\left(2+\sqrt{3}\right)\tau_T,\label{t24}
\end{equation}
which are equivalent with
\begin{equation}
\left(2-\sqrt{3}\right)\tau_q<\tau_T<\left(2+\sqrt{3}\right)\tau_q.\label{t25}
\end{equation}

\section{Case $(n,m)\in \{(3,0),(0,3)\}$}
Let us first consider the constitutive equation
\begin{equation}
q_i(t)=-k_{ij}\left[T_{,j}(t)+\tau_T\dot T_{,j}(t)+\frac12\, \tau_T^2\ddot T_{,j}(t)+\frac16\, \tau_T^3 \dddot T_{,j}(t)\right],\label{t26}
\end{equation}
and note that for any cycle as described by (\ref{t9}) we have
\begin{equation}
\int_0^{2\pi/\omega} q_i(t)T_{,i}(t)dt=-\frac{\pi}{2\omega}\left(k_{ij}f_if_j+k_{ij}g_ig_j\right)
\left(2-\tau_T^2\omega^2\right),\label{t27}
\end{equation}
which cannot conserve a constant sign for all $\omega>0$. Thus, the Second Law of Thermodynamics cannot be satisfied and the corresponding dual-phase-lag model is inconsistently from thermodynamic point of view.

Let us now consider the constitutive equation
\begin{equation}
q_i(t)+\tau_q\dot q_i(t)+\frac12\, \tau_q^2\ddot q_i(t)+\frac16\, \tau_q^3\dddot q_i(t)=-k_{ij}T_{,j}(t),\label{t28}
\end{equation}
which can be written as
\begin{equation}
T_{,i}(t)=-K_{ij}\left[q_j(t)+\tau_q\dot q_j(t)+\frac12\, \tau_q^2\ddot q_j(t)+\frac16\, \tau_q^3\dddot q_j(t)\right].\label{t29}
\end{equation}
Thus, for any cycle characterized by
\begin{equation}
q_i(t-s)=h_i\cos\omega (t-s)+\ell_i\sin\omega (t-s), \ \omega>0, \ h_ih_i+\ell_i\ell_i>0,\label{t30}
\end{equation}
we have
\begin{equation}
\int_0^{2\pi/\omega} q_i(t)T_{,i}(t)dt=-\frac{\pi\left(K_{ij}h_ih_j+K_{ij}\ell_i\ell_j\right)}{2\omega}
\left(2-\tau_q^2\omega^2\right),\label{t31}
\end{equation}
which cannot conserve a constant sign for all $\omega>0$. Thus, the Second Law of Thermodynamics cannot be satisfied and the corresponding dual-phase-lag model is inconsistently from thermodynamic point of view.

\section{Case $(n,m)\in \{(3,1),(1,3)\}$}

We consider first the constitutive equation
\begin{equation}
q_i(t)+\tau_q \dot q_i(t)=-k_{ij}\left[T_{,j}(t)+\tau_T\dot T_{,j}(t)+\frac12\, \tau_T^2\ddot T_{,j}(t)+\frac16\, \tau_T^3 \dddot T_{,j}(t)\right],\label{t32}
\end{equation}
from which we deduce the following representation
\begin{equation}
q_i(t)=-\frac{1}{\tau_q}\int_0^{\infty}e^{-s/\tau_q}k_{ij}\left[T_{,j}(t-s)+\tau_T\dot T_{,j}(t-s)+\frac12\, \tau_T^2\ddot T_{,j}(t-s)+\frac16\, \tau_T^3 \dddot T_{,j}(t-s)\right]ds.\label{t33}
\end{equation}
Further, for any cycle as described by (\ref{t9}), we obtain
\begin{equation}
\begin{split}
\int_0^{2\pi/\omega}q_i(t)T_{,i}(t)dt= \frac{\pi\left(k_{ij}f_if_j+k_{ij}g_ig_j\right)}{\omega\left(1+\tau_q^2\omega^2\right)}
\left[\frac16\tau_q\tau_T^3\omega^4+\left(\frac12\, \tau_T^2-\tau_q\tau_T\right)\omega^2-1\right],\label{t34}
\end{split}
\end{equation}
an expression which cannot be negative for all positive values of $\omega$. Thus, the dual-phase-lag model based on the constitutive equation (\ref{t32}) cannot be compatible with thermodynamics.

When the case $(n,m)=(3,1)$ is addressed we write the constitutive equation in the form
\begin{equation}
T_{,i}(t)+\tau_T \dot T_{,i}(t)=-K_{ij}\left[q_{j}(t)+\tau_q\dot q_{j}(t)+\frac12\, \tau_q^2\ddot q_{j}(t)+\frac16\, \tau_q^3 \dddot q_{j}(t)\right],\label{t35}
\end{equation}
and use the above procedure to establish the incompatibility with thermodynamics of the corresponding dual-phase-lag model of heat conduction.

\section{Case $(n,m)\in \{(3,2),(2,3)\}$}

Let us first consider the case $(n,m)=(2,3)$, that is we study the constitutive equation
\begin{equation}
q_i(t)+\tau_q \dot q_i(t)+\frac12\, \tau_q^2 \ddot q_i(t)=-k_{ij}\left[T_{,j}(t)+\tau_T\dot T_{,j}(t)+\frac12\, \tau_T^2\ddot T_{,j}(t)+\frac16\, \tau_T^3 \dddot T_{,j}(t)\right].\label{t36}
\end{equation}
It is equivalent with the following representation
\begin{equation}
q_i(t)=-\frac{2}{\tau_q}\int_0^{\infty}e^{-s/\tau_q}\left(\sin \frac{s}{\tau_q}\right)\, k_{ij}\left[T_{,j}(t-s)+\tau_T\dot T_{,j}(t-s)+\frac12\, \tau_T^2\ddot T_{,j}(t-s)+\frac16\, \tau_T^3 \dddot T_{,j}(t-s)\right]ds,\label{t37}
\end{equation}
and therefore, for any cycle as described by (\ref{t9}), we obtain
\begin{equation}
\begin{split}
\int_0^{2\pi/\omega}q_i(t)T_{,i}(t)dt=&-\frac{2\pi\left(k_{ij}f_if_j+k_{ij}g_ig_j\right)}
{\omega(\tau_q^4\omega^4+4)}\left[\frac12\, \frac{\tau_T^2}{\tau_q^2}\left(1-\frac23 \, \frac{\tau_T}{\tau_q}\right)\tau_q^4\omega^4\right.\\
&\left.+\left(2\frac{\tau_T}{\tau_q}-1-\frac{\tau_T^2}{\tau_q^2}\right)\tau_q^2\omega^2+2\right],\label{t38}
\end{split}
\end{equation}
which is negative for all positive $\omega$ if the delay times satisfy the inequality
\begin{equation}
0.28441\, \tau_q<\tau_T\leq 1.4902\, \tau_q.\label{t39}
\end{equation}

In the other case, that is for $(n,m)=(3,2)$, we write the constitutive equation in the following form
\begin{equation}
T_{,i}(t)+\tau_T \dot T_{,i}(t)+\frac12\, \tau_T^2 \ddot T_{,i}(t)=-K_{ij}\left[q_{j}(t)+\tau_q\dot q_{j}(t)+\frac12\, \tau_q^2\ddot q_{j}(t)+\frac16\, \tau_q^3 \dddot q_{j}(t)\right],\label{t40}
\end{equation}
and use the same procedure like that in the above to obtain the following restriction
\begin{equation}
0.28441\, \tau_T<\tau_q\leq 1.4902\, \tau_T.\label{t41}
\end{equation}

Thus, we can conclude that the restrictions to be fulfilled by the delay times in order to have the thermodynamic consistency are described by the relations (\ref{t39}) and (\ref{t41}), respectively.

\section{Case $(n,m)=(3,3)$}

We consider here the constitutive equation
\begin{equation}
q_i(t)+\tau_q\dot q_i(t)+\frac{\tau_q^2}{2}\, \ddot q_i(t)+\frac{\tau_q^3}{6}\, \dddot q_i(t)=-k_{ij}\left[T_{,j}(t)+\tau_T\dot T_{,j}(t)+\frac{\tau_T^2}{2}\ddot T_{,j}(t)+\frac{\tau_T^3}{6}\, \dddot T_{,j}(t)\right],\label{t42}
\end{equation}
from which we deduce
\begin{equation}
\begin{split}
q_i(t)=&-\frac{6}{\tau_q\left[\left(\alpha-\gamma\right)^2+\delta^2\right]}\int_0^{\infty}\kappa(s)k_{ij}\left[T_{,j}(t-s)+\tau_T\dot T_{,j}(t-s)\right.\\
&\left. +\frac{\tau_T^2}{2}\, \ddot T_{,j}(t-s)+\frac{\tau_T^3}{6}\, \dddot T_{,j}(t-s)\right]ds,\label{t43}
\end{split}
\end{equation}
where, for convenience, we have set
\begin{equation}
\kappa(s)=e^{-\alpha s/\tau_q}+e^{-\gamma s/\tau_q}\left(\frac{\alpha -\gamma}{\delta}\sin\frac{\delta s}{\tau_q}-\cos\frac{\delta s}{\tau_q}\right),\label{t44}
\end{equation}
and
\begin{equation}
\alpha=1.5961,\quad \gamma=0.70196,\quad \delta=1.8073.\label{t45}
\end{equation}
Then, for any cycle characterized by (\ref{t9}), we have
\begin{equation}
\begin{split}
\int_0^{2\pi/\omega}q_i(t)T_{,i}(t)dt=-\frac{6\pi\left(k_{ij}f_if_j+k_{ij}g_ig_j\right)}
{\tau_q\omega\left[(\alpha-\gamma)^2+\delta^2\right]}\left[\left(1-\frac12\, \tau_T^2\omega^2\right)\kappa_c+\left(\tau_T\omega-\frac16\, \tau_T^3\omega^3\right)\kappa_s\right],\label{t46}
\end{split}
\end{equation}
where
\begin{equation}
\kappa_c=\int_0^{\infty}\kappa(s)\cos \omega sds, \quad \kappa_s=\int_0^{\infty}\kappa(s)\sin \omega sds.\label{t47}
\end{equation}

Furthermore, from the relations (\ref{t44}) and (\ref{t47}), we have
\begin{equation}
\begin{split}
\kappa_c=&\tau_q\left[\frac{\alpha}{\alpha^2+\tau_q^2\omega^2}+\frac{\left(\alpha-2\gamma\right)
\left(\gamma^2+\delta^2\right)-\alpha\tau_q^2\omega^2}
{\left(\gamma^2+\delta^2+\tau_q^2\omega^2\right)^2-4\delta^2\tau_q^2\omega^2}\right],\\
\kappa_s=&\tau_q^2\omega\left[\frac{1}{\alpha^2+\tau_q^2\omega^2}+\frac{\delta^2-3\gamma^2+2\gamma\alpha-\tau_q^2\omega^2}
{\left(\gamma^2+\delta^2+\tau_q^2\omega^2\right)^2-4\delta^2\tau_q^2\omega^2}\right].\label{t48}
\end{split}
\end{equation}

In view of the values for $\alpha$, $\gamma$ and $\delta$ given in (\ref{t45}), it follows that the constitutive equation (\ref{t42}) is compatible with thermodynamics if the following condition is fulfilled
\begin{equation}
\begin{split}
&0.677637\, \frac{\tau_T^3}{\tau_q^3}\left(\tau_q^2\omega^2\right)^3-\frac{\tau_T}{\tau_q}\left(4.0657\, \frac{\tau_T^2}{\tau_q^2}-6.09877\, \frac{\tau_T}{\tau_q}+4.06582 \right)\left(\tau_q^2\omega^2\right)^2\\
&-\left[12.1972\, \left(\frac{\tau_T}{\tau_q}-1\right)^2+0.0003\right]\left(\tau_q^2\omega^2\right)+24.3944\geq 0,\label{t49}
\end{split}
\end{equation}
for all $\omega \geq 0$. We associate with this inequality the function
\begin{equation}
f(z)\equiv a_1z^3-b_1z^2-c_1z+d_1,\label{t491}
\end{equation}
where
\begin{equation}
\begin{split}
a_1=&0.677637\, \frac{\tau_T^3}{\tau_q^3},\quad b_1=\frac{\tau_T}{\tau_q}\left(4.06574\, \frac{\tau_T^2}{\tau_q^2}-6.09877\, \frac{\tau_T}{\tau_q}+4.06582 \right)\\
c_1=&12.1972\, \left(\frac{\tau_T}{\tau_q}-1\right)^2+0.0003,\quad d_1=24.3944.\label{t51}
\end{split}
\end{equation}
Furthermore, we see that the derivative
\begin{equation}
\frac{df}{dz}=3a_1z^2-2b_1z-c_1,\label{t492}
\end{equation}
admits the roots
\begin{equation}
z_1=\frac{b_1-\sqrt{b_1^2+3a_1c_1}}{3a_1}<0,\quad z_2=\frac{b_1+\sqrt{b_1^2+3a_1c_1}}{3a_1}>0.\label{t493}
\end{equation}
Thus, $f(z)\geq 0$ for all $z>0$ if we have $f(z_2)>0$, that is
\begin{equation}
\sqrt{\left(b_1^2+3a_1c_1\right)^3}\leq \frac12\, \left(27a_1^2d_1-2b_1^3-9a_1b_1c_1\right).\label{t50}
\end{equation}

Thus, the constitutive equation (\ref{t42}) is compatible with the thermodynamics if the delay times satisfy the inequality (\ref{t50}). It can be verified that the set of delay times fulfilling (\ref{t50}) is non-empty (in fact it contains the case when $\tau_T=\tau_q$). Moreover, we have to outline that the inequality (\ref{t49}) has an approximately character and it depends on the approximation order of the roots $\alpha$, $\gamma$ and $\delta$ as defined in (\ref{t45}) and obtained in the Table 1.

\section{Case $(n,m)\in \{(4,0),(0,4)\}$}
Let us first consider the constitutive equation
\begin{equation}
q_i(t)=-k_{ij}\left[T_{,j}(t)+\tau_T\dot T_{,j}(t)+\frac12\, \tau_T^2\ddot T_{,j}(t)+\frac16\, \tau_T^3 \dddot T_{,j}(t)+\frac{1}{24}\, \tau_T^4\, \frac{\partial^4 T_{,j}}{\partial t^4}(t)\right],\label{t52}
\end{equation}
and note that for any cycle as described by (\ref{t9}) we have
\begin{equation}
\int_0^{2\pi/\omega} q_i(t)T_{,i}(t)dt=-\frac{\pi\left(k_{ij}f_if_j+k_{ij}g_ig_j\right)}{24\omega}
\left[\left(\tau_T^2\omega^2-6\right)^2-12\right],\label{t53}
\end{equation}
which cannot conserve a constant sign for all $\omega>0$. Thus, the Second Law of Thermodynamics cannot be satisfied and the corresponding dual-phase-lag model is inconsistently from thermodynamic point of view.

Let us now consider the constitutive equation
\begin{equation}
q_i(t)+\tau_q\dot q_i(t)+\frac12\, \tau_q^2\ddot q_i(t)+\frac16\, \tau_q^3\dddot q_i(t)+\frac{1}{24}\, \tau_q^4\, \frac{\partial^4 q_i}{\partial t^4}(t)=-k_{ij}T_{,j}(t),\label{t54}
\end{equation}
which can be written as
\begin{equation}
T_{,i}(t)=-K_{ij}\left[q_j(t)+\tau_q\dot q_j(t)+\frac12\, \tau_q^2\ddot q_j(t)+\frac16\, \tau_q^3\dddot q_j(t)+\frac{1}{24}\, \tau_q^4\, \frac{\partial^4 q_j}{\partial t^4}(t)\right].\label{t55}
\end{equation}
Thus, for any cycle characterized by (\ref{t30}), we obtain
\begin{equation}
\int_0^{2\pi/\omega} q_i(t)T_{,i}(t)dt=-\frac{\pi\left(K_{ij}h_ih_j+K_{ij}\ell_i\ell_j\right)}{24\omega}
\left[\left(\tau_q^2\omega^2-6\right)^2-12\right],\label{t56}
\end{equation}
which cannot conserve a constant sign for all $\omega>0$. Thus, the Second Law of Thermodynamics cannot be satisfied and the corresponding dual-phase-lag model is inconsistently from thermodynamic point of view.

\section{Case $(n,m)\in \{(4,1),(1,4)\}$}
We consider now the constitutive equation
\begin{equation}
q_i(t)+\tau_q\dot q_i(t)=-k_{ij}\left[T_{,j}(t)+\tau_T\dot T_{,j}(t)+\frac12\, \tau_T^2\ddot T_{,j}(t)+\frac16\, \tau_T^3 \dddot T_{,j}(t)+\frac{1}{24}\, \tau_T^4\, \frac{\partial^4 T}{\partial t^4}(t)\right],\label{t57}
\end{equation}
which furnishes
\begin{equation}
\begin{split}
q_i(t)=&-\frac{1}{\tau_q}\, \int_0^{\infty}e^{-s/\tau_q}\, k_{ij}\left[T_{,j}(t-s)+\tau_T\dot T_{,j}(t-s)+\frac12\, \tau_T^2 \ddot T_{,j}(t-s)\right.\\
&\left. +\frac16\, \tau_T^3  \dddot T_{,j}(t-s)+\frac{1}{24}\, \tau_T^4\, \frac{\partial^4 T}{\partial t^4}(t-s)\right]ds.\label{t58}
\end{split}
\end{equation}
Therefore, for any cycle characterized by (\ref{t9}), we get
\begin{equation}
\begin{split}
\int_0^{2\pi/\omega} q_i(t)T_{,i}(t)dt=&-\frac{\pi\left(k_{ij}f_if_j+k_{ij}g_ig_j\right)}{\omega\left(1+\tau_q^2\omega^2\right)}\\
&\times\left[\frac{\tau_T^3}{6\tau_q^3}\left(\frac{\tau_T}{4\tau_q}-1\right)\tau_q^4\omega^4+
\frac{\tau_T}{\tau_q}\left(1-\frac{\tau_T}{2\tau_q}\right)\tau_q^2\omega^2+1\right],\label{t59}
\end{split}
\end{equation}
which cannot conserve a constant sign for all $\omega>0$. Thus, the Second Law of Thermodynamics cannot be satisfied and the corresponding dual-phase-lag model is inconsistently from thermodynamic point of view.

When it is considered the constitutive equation
\begin{equation}
q_i(t)+\tau_q\dot q_i(t)+\frac12\, \tau_q^2\ddot q_i(t)+\frac16\, \tau_q^3\dddot q_i(t)+\frac{1}{24}\, \tau_q^4\frac{\partial^4 q_i}{\partial t^4}(t)=-k_{ij}\left[T_{,j}(t)+\tau_T\dot T_{,j}(t)\right],\label{t60}
\end{equation}
we write it as
\begin{equation}
T_{,i}(t)+\tau_T\dot T_{,i}(t)=-K_{ij}\left[q_j(t)+\tau_q\dot q_j(t)+\frac12\, \tau_q^2\ddot q_j(t)+\frac16\, \tau_q^3\dddot q_j(t)+\frac{1}{24}\, \tau_q^4\frac{\partial^4 q_j}{\partial t^4}(t)\right],\label{t61}
\end{equation}
and we obtain
\begin{equation}
\begin{split}
\int_0^{2\pi/\omega} q_i(t)T_{,i}(t)dt=&-\frac{\pi\left(K_{ij}h_ih_j+K_{ij}\ell_i\ell_j\right)}{\omega\left(1+\tau_T^2\omega^2\right)}\\
&\times\left[\frac{\tau_q^3}{6\tau_T^3}
\left(\frac{\tau_q}{4\tau_T}-1\right)\tau_T^4\omega^4+\frac{\tau_q}{\tau_T}\left(1-\frac{\tau_q}{2\tau_T}\right)
\tau_T^2\omega^2+1\right],\label{t62}
\end{split}
\end{equation}
which cannot conserve a constant sign for all $\omega>0$.

Concluding this section, we see that the dual-phase-lag models based on the constitutive equations (\ref{t57}) and (\ref{t60})  are not compatible with the thermodynamics.

\section{Case $(n,m)\in \{(4,2),(2,4)\}$}

Let us first consider the constitutive equation
\begin{equation}
\begin{split}
q_i(t)+\tau_q \dot q_i(t)+\frac12\, \tau_q^2\ddot q_i(t)=&-k_{ij}\left[T_{,j}(t)+\tau_T\dot T_{,j}(t)+\frac12\, \tau_T^2\ddot T_{,j}(t)\right.\\
& \left.+\frac16\, \tau_T^3\dddot T_{,j}(t)+\frac{1}{24}\, \tau_T^4\, \frac{\partial^4 T_{,j}}{\partial t^4}(t)\right],\label{t63}
\end{split}
\end{equation}
which gives
\begin{equation}
\begin{split}
q_i(t)=&-\frac{2}{\tau_q}\int_0^{\infty}e^{-s/\tau_q}\sin\frac{s}{\tau_q}k_{ij}\left[T_{,j}(t-s)+\tau_T\dot T_{,j}(t-s)+\frac12\, \tau_T^2\ddot T_{,j}(t-s)\right.\\
&\left.+\frac16\, \tau_T^3\dddot T_{,j}(t-s)+\frac{1}{24}\, \tau_T^4\, \frac{\partial^4 T_{,j}}{\partial t^4}(t-s)\right]ds.\label{t64}
\end{split}
\end{equation}
Further, for any cycle characterized by (\ref{t9}), we obtain
\begin{equation}
\begin{split}
\int_0^{2\pi/\omega}q_i(t)T_{,i}(t)dt=&-\frac{\pi\left(k_{ij}f_if_j+k_{ij}g_ig_j\right)}
{\omega(\tau_q^4\omega^4+4)}\left[-\frac{\tau_T^4}{12\tau_q^4}\left(\tau_q^6\omega^6\right)\right.\\
&\left.+\left(\frac{\tau_T^4}{6\tau_q^4}-\frac{2\tau_T^3}{3\tau_q^3}+ \frac{\tau_T^2}{\tau_q^2}\right)\left(\tau_q^4\omega^4\right)-
2\left(1-\frac{2\tau_T}{\tau_q}+\frac{\tau_T^2}{\tau_q^2}\right)\left(\tau_q^2\omega^2\right)+4\right],\label{t64}
\end{split}
\end{equation}
which cannot conserve a negative sign for all $\omega>0$. Thus, the Second Law of Thermodynamics cannot be satisfied and the corresponding dual-phase-lag model is inconsistently from thermodynamic point of view. The same conclusion can be obtained for the case $(n,m)=(4,2)$.

\section{Case $(n,m)\in \{(4,3),(3,4)\}$}

Let us first consider the constitutive equation
\begin{equation}
\begin{split}
&q_i(t)+\tau_q \dot q_i(t)+\frac12\, \tau_q^2\ddot q_i(t)+\frac16\, \tau_q^3\dddot q_i(t)=-k_{ij}\left[T_{,j}(t)+\tau_T\dot T_{,j}(t)+\frac12\, \tau_T^2\ddot T_{,j}(t)\right.\\
& \left.+\frac16\, \tau_T^3\dddot T_{,j}(t)+\frac{1}{24}\, \tau_T^4\, \frac{\partial^4 T_{,j}}{\partial t^4}(t)\right],\label{t66}
\end{split}
\end{equation}
which gives
\begin{equation}
\begin{split}
q_i(t)=&-\frac{6}{\tau_q\left[\left(\gamma-\alpha\right)^2+\delta^2\right]}\int_0^{\infty}\kappa(s)k_{ij}
\left[T_{,j}(t-s)+\tau_T\dot T_{,j}(t-s)\right.\\
&\left.+\frac12 \, \tau_T^2\ddot T_{,j}(t-s)+\frac16\, \tau_T^3\dddot T_{,j}(t-s)+\frac{1}{24}\, \tau_T^4\, \frac{\partial^4 T_{,j}}{\partial t^4}(t-s) \right]ds,\label{t67}
\end{split}
\end{equation}
where $\kappa(s)$ is given by relation (\ref{t44}) and the values of $\alpha$, $\delta$ and $\gamma$ are given by (\ref{t45}).
For any cycle characterized by (\ref{t9}), we have
\begin{equation}
\begin{split}
\int_0^{2\pi/\omega}q_i(t)T_{,i}(t)dt=&-\frac{6\pi\left(k_{ij}f_if_j+k_{ij}g_ig_j\right)}
{\tau_q\omega\left[\left(\gamma-\alpha\right)^2+\delta^2\right]}\left[\left(1-\frac12\, \tau_T^2\omega^2+\frac{1}{24}\, \tau_T^4\omega^4\right)\kappa_c\right.\\
&\left.+\left(\tau_T\omega-\frac16\, \tau_T^3\omega^3\right)\kappa_s\right],\label{t68}
\end{split}
\end{equation}
where $\kappa_c$ and $\kappa_s$ are defined by the relation (\ref{t48}).
In view of relations (\ref{t45}), (\ref{t48}) and (\ref{t68}) it follows that the Second Law of Thermodynamics can be satisfied if the following inequality is fulfilled
\begin{equation}
\begin{split}
&\left(0.677637\, \frac{\tau_T^3}{\tau_q^3}-0.508231\,\frac{\tau_T^4}{\tau_q^4}\right)\left(\tau_q^2\omega^2\right)^3+
\frac{\tau_T}{\tau_q}\left(1.01643\, \frac{\tau_T^3}{\tau_q^3}- 4.06574\, \frac{\tau_T^2}{\tau_q^2}\right.\\
&\left.+6.09877\, \frac{\tau_T}{\tau_q}-4.06582 \right)\left(\tau_q^2\omega^2\right)^2\\
&-\left[12.1972\, \left(\frac{\tau_T}{\tau_q}-1\right)^2+0.0003\right]\left(\tau_q^2\omega^2\right)+24.3944\geq 0,\label{t69}
\end{split}
\end{equation}
for all $\omega \geq 0$. Thus, the constitutive equation (\ref{t66}) is compatible with the thermodynamics if the delay times satisfy the inequalities
\begin{equation}
\begin{split}
&0<\frac{\tau_T}{\tau_q}<1.33332,\\
&\sqrt{\left(b_2^2+3a_2c_2\right)^3}\leq \frac12\, \left(27a_2^2d_2-2b_2^3-9a_2b_2c_2\right),\label{t70}
\end{split}
\end{equation}
where
\begin{equation}
\begin{split}
a_2=&0.677637\, \frac{\tau_T^3}{\tau_q^3}-0.508231\, \frac{\tau_T^4}{\tau_q^4} \\
b_2=&\frac{\tau_T}{\tau_q}\left(-1.01643\, \frac{\tau_T^3}{\tau_q^3}+4.06574\, \frac{\tau_T^2}{\tau_q^2}-6.09877\, \frac{\tau_T}{\tau_q}+4.06582 \right)\\
c_2=&12.1972\, \left(\frac{\tau_T}{\tau_q}-1\right)^2+0.0003,\\
d_2=&24.3944.\label{t71}
\end{split}
\end{equation}

It can be verified that it is non-empty the set of values for the delay times as defined by the relation (\ref{t70}), (in fact it contains at least the case when the delay times are equally, $\tau_T=\tau_q$).

When the case $(n,m)=(4,3)$ is considered, the constitutive equation is written in the equivalent form
\begin{equation}
\begin{split}
&T_{,i}(t)+\tau_T \dot T_{,i}(t)+\frac12\, \tau_T^2\ddot T_{,i}(t)+\frac16\, \tau_T^3\dddot T_{,i}(t)=-K_{ij}\left[q_{j}(t)+\tau_q\dot q_{j}(t)+\frac12\, \tau_q^2\ddot q_{j}(t)\right.\\
& \left.+\frac16\, \tau_q^3\dddot q_{j}(t)+\frac{1}{24}\, \tau_q^4\, \frac{\partial^4 q_{j}}{\partial t^4}(t)\right],\label{t72}
\end{split}
\end{equation}
and then the thermodynamic restrictions can be obtained easily by using the relation (\ref{t70}).

\section{Case $(n,m)=(4,4)$}

Let us now consider the constitutive equation
\begin{equation}
\begin{split}
&q_i(t)+\tau_q\dot q_i(t)+\frac12\, \tau_q^2\ddot q_i(t)+\frac16\, \tau_q^3\dddot q_i(t)+\frac{1}{24}\, \tau_q^4\, \frac{\partial^4 q_i}{\partial t^4}(t)\\
&=-k_{ij}\left[T_{,j}(t)+\tau_T\dot T_{,j}(t)+\frac12\, \tau_T^2\ddot T_{,j}(t)+\frac16\, \tau_T^3\dddot T_{,j}(t)+\frac{1}{24}\, \tau_T^4\, \frac{\partial^4 T_{,j}}{\partial t^4}(t)\right],\label{t76}
\end{split}
\end{equation}
which can be written as
\begin{equation}
\begin{split}
q_i(t)=&-\frac{24}{\tau_q\Delta}\int_0^{\infty}K(s)k_{ij}\left[T_{,j}(t-s)+\tau_T\dot T_{,j}(t-s)\right.\\
&\left. +\frac{\tau_T^2}{2}\, \ddot T_{,j}(t-s)+\frac{\tau_T^3}{6}\, \dddot T_{,j}(t-s)+\frac{1}{24}\, \tau_T^4\, \frac{\partial^4 T_{,j}}{\partial t^4}(t-s)\right]ds,\label{t77}
\end{split}
\end{equation}
where
\begin{equation}
\begin{split}
K(s)=&e^{-\alpha s/\tau_q}\left[\cos \frac{\beta s}{\tau_q}-\frac{(\alpha-\gamma)^2+\delta^2-\beta^2}{2\beta(\gamma-\alpha)}\, \sin \frac{\beta s}{\tau_q}\right]\\
&-e^{-\gamma s/\tau_q}\left[\cos \frac{\delta s}{\tau_q}+\frac{(\alpha-\gamma)^2+\beta^2-\delta^2}{2\delta(\gamma-\alpha)}\, \sin \frac{\delta s}{\tau_q}\right],\label{t78}
\end{split}
\end{equation}
\begin{equation}
\begin{split}
\Delta=&3\alpha\beta^2-3\gamma\delta^2+\gamma^3-\alpha^3-\frac{1}{2(\gamma-\alpha)}
\left[\left(\gamma-\alpha\right)^2\left(3\alpha^2+3\gamma^2-\beta^2-\delta^2\right)\right.\\
&\left.+\left(\beta^2-\delta^2\right)
\left(3\gamma^2-3\alpha^2+\beta^2-\delta^2\right)\right]=-22.165,\label{t79}
\end{split}
\end{equation}
and now we have set
\begin{equation}
\begin{split}
\alpha=&0.27056,\quad \beta=2.5048,\\
\gamma=&1.7294,\quad \delta=0.88897.\label{t80}
\end{split}
\end{equation}
Then, for any cycle characterized by (\ref{t9}), we have
\begin{equation}
\begin{split}
\int_0^{2\pi/\omega}q_i(t)T_{,i}(t)dt=&-\frac{24\pi\left(k_{ij}f_if_j+k_{ij}g_ig_j\right)}
{\tau_q\omega\Delta }\left[\left(1-\frac12\, \tau_T^2\omega^2+\frac{1}{24}\, \tau_T^4\omega^4\right)K_c\right.\\
+&\left.\left(\tau_T\omega-\frac16\, \tau_T^3\omega^3\right)K_s\right],\label{t81}
\end{split}
\end{equation}
where
\begin{equation}
K_c=\int_0^{\infty}K(s)\cos \omega sds,\quad K_s=\int_0^{\infty}K(s)\sin \omega sds.\label{t82}
\end{equation}

Further, with the aid of (\ref{t78}), we have
\begin{equation}
\begin{split}
K_c=&\frac{\tau_q}{2(\gamma-\alpha)}\, \left[ \frac{\left(\gamma^2+\delta^2-\alpha^2-\beta^2\right)\tau_q^2\omega^2-\left(\alpha^2+\beta^2\right)
\left(3\alpha^2-\beta^2+\gamma^2+\delta^2-4\alpha\gamma\right)}{\tau_q^4\omega^4+2(\alpha^2-\beta^2)\tau_q^2\omega^2+\left(\alpha^2+\beta^2\right)^2}\right.\\
&\left.-\frac{\left(\gamma^2+\delta^2-\alpha^2-\beta^2\right)\tau_q^2\omega^2+\left(\gamma^2+\delta^2\right)
\left(3\gamma^2-\delta^2+\alpha^2+\beta^2-4\alpha\gamma\right)}{\tau_q^4\omega^4+2(\gamma^2-\delta^2)\tau_q^2\omega^2+\left(\gamma^2+\delta^2\right)^2}\right]
\label{t83}
\end{split}
\end{equation}
and
\begin{equation}
\begin{split}
K_s=\tau_q^2\omega & \left[\frac{\tau_q^2\omega^2+\alpha^2-\beta^2-\frac{\alpha}{\gamma-\alpha}
\left[(\gamma-\alpha)^2+\delta^2-\beta^2\right]}{\tau_q^4\omega^4+2\left(\alpha^2-\beta^2\right)\tau_q^2\omega^2+\left(\alpha^2+\beta^2\right)^2}
\right.\\
-&\left.\frac{\tau_q^2\omega^2+\gamma^2-\delta^2+
\frac{\gamma}{\gamma-\alpha}
\left[(\gamma-\alpha)^2+\beta^2-\delta^2\right]}{\tau_q^4\omega^4+2\left(\gamma^2-\delta^2\right)\tau_q^2\omega^2+\left(\gamma^2+\delta^2\right)^2}\right].\label{t84}
\end{split}
\end{equation}
In view of relations (\ref{t80}), (\ref{t83}) and (\ref{t84}) it follows that the Second Law of Thermodynamics can be satisfied if the following inequality is fulfilled
\begin{equation}
\begin{split}
&2.69456\,\frac{\tau_T^4}{\tau_q^4}\left(\tau_q^2\omega^2\right)^4-
\frac{\tau_T^2}{\tau_q^2}\left(32.3346\, \frac{\tau_T^2}{\tau_q^2}- 43.1121\, \frac{\tau_T}{\tau_q}+32.3348\right)\left(\tau_q^2\omega^2\right)^3\\
&+\left(64.668\, \frac{\tau_T^4}{\tau_q^4}-258.676\, \frac{\tau_T^3}{\tau_q^3} +388.015\, \frac{\tau_T^2}{\tau_q^2}-258.673\, \frac{\tau_T}{\tau_q}+64.6695 \right)\left(\tau_q^2\omega^2\right)^2\\
&-\left(776.016\, \frac{\tau_T^2}{\tau_q^2}-1552.06\, \frac{\tau_T}{\tau_q}+776.031\right)\left(\tau_q^2\omega^2\right)+1552.03\geq 0,\label{t85}
\end{split}
\end{equation}
for all $\omega \geq 0$. It is not difficult to see that the above inequality is fulfilled when $\tau_T=\tau_q$ and so we can conclude that it is non-empty the set of couples $(\tau_T,\tau_q)$ for which the constitutive equation (\ref{t76}) is compatible with the thermodynamics. Unfortunately, there seems to be unavailable to get an explicit expression defining this set.

\section{Conclusions}

We inferred here an opinion about the time differential dual-phase-lag models that is based on the information that we have upon the differential operators involved into the related constitutive equations. It is shown that, when the approximation orders are greater than or equal to five, the corresponding models lead to some instable systems. Instead, when the approximation orders are lower than or equal to four, then the corresponding models can be compatible with the thermodynamics, provided some appropriate restrictions are assumed upon the delay times. More precisely, the thermodynamic consistency of the model in concern is established when $(m,n)\in \{(0,0), (1,0), (0,1), (2,1), (1,2), (2,2), (3,2), (2,3), (3,3), (3,4), (4,3), (4,4)\}$, provided appropriate restrictions are placed on the delay times.

On the other side, the present paper furnishes a new perspective upon the relationship between the thermodynamic aspects and the stability properties, with reference to the time differential dual-phase-lag models of heat conduction. We believe that this research may be useful to clarify the
applicability of these theories.

\section*{References}

\section*{Appendix: Szeg\"{o}'s curve}

In this appendix we provide the source code that has been used to generate an animation with the software package Wolfram Mathematica 11.
The generated animation shows the complex roots of the scaled exponential sum of order $n$ in the complex plane, with $n$ increasing from $1$ to $50$, and it shows that the roots approach the Szeg\"o curve as $n$ increases. 

The following code could be entered in a Mathematica session and, once executed, the animation window is shown.

\begin{verbatim}
e[n_, z_] := Sum[z^k/k!, {k, 0, n}]
axes = Graphics[{Line[{{-1, 0}, {1, 0}}], Line[{{0, -1}, {0, 1}}]}];
szegoeCurve = ContourPlot[Evaluate[Abs[z E^(1 - z)] == 1 /. z -> x + I y],
  {x, -1, 1}, {y, -1, 1}, PlotPoints -> 50];
rootsGraph[m_] := Module[{roots, n = IntegerPart[m], points},
  roots = NSolve[e[n, n z] == 0, z][[All, 1, 2]];
  points = Transpose[{Re[roots], Im[roots]}];
  Show[{axes, szegoeCurve, ListPlot[points,
    PlotRange -> {{-1, 1}, {-1, 1}}, AspectRatio -> 1]}]]
Animate[rootsGraph[m], {m, 1, 50}, DefaultDuration -> 50]
\end{verbatim}

Here we also provide an image of the roots for $n=25$:

\end{document}